# Weyl Fermion Magneto-Electrodynamics and Ultra-low Field Quantum Limit in TaAs

Zhengguang Lu[1], Patrick Hollister[2], Mykhaylo Ozerov[3], Seongphill Moon[3,4], Eric D. Bauer[5], Filip Ronning[5], Dmitry Smirnov[3], Long Ju[1*], B. J. Ramshaw[2*]

1. Department of physics, Massachusetts Institute of Technology, Cambridge, MA 02139, USA
2. Laboratory of Atomic and Solid State Physics, Cornell University, Ithaca, NY 14853, USA.
3. National High Magnetic Field Lab, Tallahassee, FL, 32310, USA
4. Department of Physics, Florida State University, Tallahassee, Florida 32306, USA
5. Los Alamos National Labs, Los Alamos, NM 87544, USA

[*] Corresponding author: bradramshaw@cornell.edu, longju@mit.edu

**Topological semimetals are predicted to exhibit unconventional electrodynamics, but a central experimental challenge is singling out the contributions from the topological bands. TaAs is the prototypical example, where 24 Weyl points and 8 trivial Fermi surfaces make the interpretation of any experiment in terms of band topology ambiguous. We report magneto-infrared reflection spectroscopy measurements on TaAs. We observed sharp inter-Landau level transitions from a single pocket of Weyl Fermions in magnetic fields as low as 0.4 tesla. We determine the W2 Weyl point to be 8.3 meV below the Fermi energy, corresponding to a quantum limit—the field required to reach the lowest LL—of 0.8 Tesla—unprecedentedly low for Weyl Fermions. LL spectroscopy allows us to isolate these Weyl Fermions from all other carriers in TaAs and our result provides a new way for directly exploring the more exotic quantum phenomena in Weyl semimetals, such as the chiral anomaly.**

Introduction

Weyl semimetals are a novel class of quantum materials featuring electronic band structures with chiral Weyl Fermions *(1–3)* and arc like surface states *(4,5)*. Weyl Fermions arise at points in the Brillouin zone where conduction and valence bands have protected crossings that arise from strong spin-orbit coupling. When time reversal and inversion symmetry are preserved, these points are four-fold-degenerate and are commonly known as Dirac points. When one of these symmetries is broken, however, the band crossings are double-degenerate—these are the Weyl points. Weyl points come in pairs, with each point having a well-defined chirality, or "handedness". The electronic band structure near the Weyl points can be described using an effective model given by the Weyl equation, first introduced for chiral, massless

(linearly-dispersing) fermions in high energy physics *(6)*. The electrodynamics of Weyl Fermions can be highly non-trivial, with phenomena that have no counterpart in regular metals such as the chiral anomaly. Since their prediction and subsequent discovery by ARPES experiments *(4,5)*, Weyl semimetals have attracted a great deal of attention, both for their unusual electrodynamics and as a platform for discovering new quantum phases of mater such as monopole superconductivity *(7)*.

However, experimental characterizations of the electrodynamics of Weyl Fermions have been significantly hampered by the complicated band structures that are found in real materials, in contrast to the ideal Weyl fermions described by the Weyl equation. In addition to "Weyl pockets"—the segments of Fermi surface that surround the Weyl points when the chemical potential is not at the band-crossing point—most Weyl systems also contain trivial (non-Weyl) Fermi surfaces. Many Weyl materials have their Weyl points far from the chemical potential, making any connection between the observed behaviour and the underlying band topology dubious at best. Finding a material with the simplest band structure or using a technique that can isolate the response of a single carrier type, is therefore essential for understanding the electrodynamic properties of Weyl Fermions.

The tantalum and niobium monopnictides (TaAs, TaP, NbAs and NbP) are one of the simplest and most widely explored families of Weyl semimetals. Among these, TaAs has the simplest low energy electronic band structure *(8)*. In the Nb based compounds, there are trivial pockets very close in momentum space to the Weyl nodes, and even merged into the Fermi surface of the Weyl pocket, which makes it almost impossible to selectively probe a particular Weyl node without contributions from the trivial pocket *(8,9)*. In addition, the Weyl points in the Nb-based compounds are separated by saddle points that are only about 20 meV apart in energy, whereas this separation is around 100 meV in the Ta based compounds *(9, 10)*. This ensures that an effective two-band model captures the essential low energy physics in the Ta-based compounds *(11,12)*. Finally, the Fermi surfaces enclosing the Weyl nodes of TaAs are much smaller than they are in TaP *(9)*, making TaAs a great model system for probing the electrodynamics of Weyl Fermions.

Band structure calculations and ARPES measurements indicate that there are two sets of Weyl points in TaAs, labeled W1 and W2 in Figure 1a. The pairs of nodes with opposite chirality (4 W1 and 8 W2 pairs) are distributed near the boundaries of the first Brillouin zone and are oriented perpendicular to the tetragonal axis *(4,9)*. TaAs has three distinct sections of Fermi surface, distributed around the dashed circle—a nodal line in the limit that spin-orbit coupling is zero—in Fig. 1a. The pockets that enclose the W1 Weyl points are highly anisotropic in their dispersion, with a smaller Fermi velocity along the crystal c-axis. The pockets that enclose the W2 Weyl points are relatively isotropic in comparison. In addition to the Weyl Fermi surfaces, there are hole pockets that do not enclose any Weyl points: these are denoted as H1 *(8,9)*.

Magneto-infrared spectroscopy has played an important role in characterizing topological phases of matter *(13)*. In contrast to magneto-transport measurements, where the DC conductivity contains contributions from all Fermi surfaces superimposed on one another, infrared spectroscopy measurements can separate the responses of different Fermi surface into different frequency ranges. Previous efforts on Weyl semimetals, however, observed inter-LL transitions only in relatively high magnetic fields (> 5 T) and at high energies (> 40 meV), leaving the electrodynamics of Weyl Fermions in the immediate vicinity of the Weyl nodes unexplored. At the same time, the Faraday configuration (with magnetic field applied perpendicular to direction of Weyl point separation) used in previous measurements on TaAs mixes signals from the two different types of Weyl Fermions *(14–17)*. Thus, an unambiguous identification of the low-energy electrodynamics from a single type of Weyl Fermion has not been reported.

**Results**

We have employed magneto-infrared reflection spectroscopy in the Voigt configuration, with the magnetic field in the *ab*-plane (see Methods and Figure 1b), to examine the low energy dynamics of Weyl Fermions around the W2 Weyl point of TaAs (Figure 1a). The Voigt configuration, combined with sharp LLs due to high sample quality, allows us to observe inter-LL transitions from only the W2 Weyl pocket and at a much lower field strength and energy range than previously reported for any other Weyl semimetal.

We applied the magnetic field in the *ab*-plane so that the Fermi surface around the W1 Weyl points has a larger cross-sectional area perpendicular to the magnetic field direction than if the field was applied along the *c*-axis. This results in a smaller cyclotron energy for quasiparticles on the W1 Fermi surface, making their contribution to the optical conductivity a featureless continuum across our entire field range. In contrast, the relatively isotropic W2 Fermi surface has a much smaller Fermi surface area perpendicular to the magnetic field. This results in a larger cyclotron energy (about one order of magnitude larger than that of W1) and well-separated peaks in the optical conductivity spectrum from inter-LL transitions.

Since the polished surface of our TaAs crystals is parallel to the ab-plane, this choice of magnetic field direction corresponds to the Voigt configuration of reflection spectroscopy, as shown in Fig. 1b. Under an applied magnetic field, electron motion perpendicular to the field direction is quantized into LLs with index *N*, as shown in Fig. 1c. These LLs evolve into a series of one-dimensional bands dispersing along $k_{//B}$, which is still a good quantum number in a magnetic field. In this case, van Hove singularities (vHSs) in the density-of-states (DOS) are located at $k_{//B} = 0$ and interband optical transitions from/to these vHSs are expected to dominate the optical conductivity spectrum. In the Voigt configuration, these optical transitions obey selection rules $\Delta|N|= \pm 1$ for light polarization perpendicular to *B* and $\Delta|N|= 0$ for polarization parallel to *B* *(18)*.

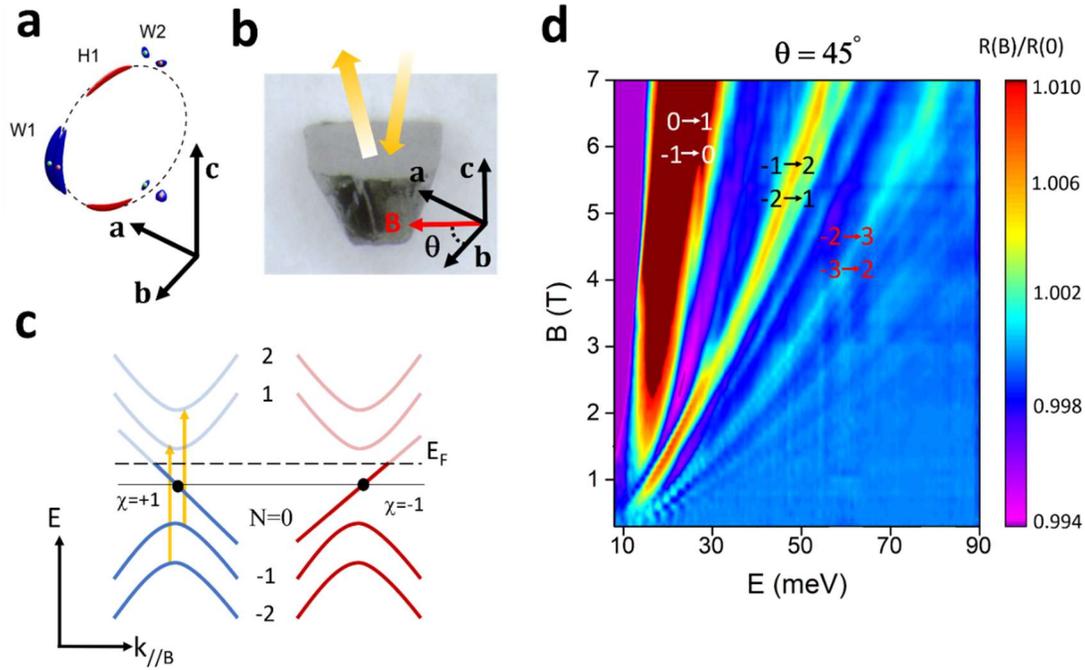

*Fig. 1 Measurement scheme and inter-LL transitions in TaAs. **a.** Fermi surface of TaAs featuring Weyl pockets W1, W2 and a trivial hole pocket H1—each distributed around the nodal line, represented by the dashed circle, that would be present with zero spin-orbit coupling (figure reproduced with permission from F. Arnold et al. (19)). There are four such nodal lines and accompanying pockets arranged at the edges of the Brillouin zone in TaAs. **b.** TaAs crystal with the flat surface in the ab-plane. The schematic indicates the measurement geometry: the magnetic field is applied in the a-b plane at an angle θ from the a-axis. Reflected infrared light is collected in the Voigt configuration. **c.** Schematic band structure for a pair of Weyl points with chirality χ=±1 in a magnetic field. LLs labeled by index N are formed in the plane perpendicular to B field, while the momentum parallel to the magnetic field, $k_{||B}$, is still a good quantum number. The zeroth LLs at both Weyl points disperse as linear, one-dimensional bands. Inter-LL transitions obey optical selection rules Δ|N|= ±1. **d.** 2D color plot of the infrared reflectivity in a magnetic field, normalized by the reflectivity at zero field (taken at θ=45°). Sharp peaks that evolve with magnetic field are labeled by the proposed LL index of initial and final states. Due to the diverging density-of-states at the band edge for each of the N>0 LLs, peaks in **d** are dominated by inter-LL transitions at $k_{||B}=\pm k_w$ where the Weyl points are located.*

Figure 1d shows the experimental reflection spectrum as a function of magnetic field. At zero magnetic field, the reflection spectrum of TaAs is expected to be smooth due to the absence of LLs and their accompanying vHSs. We focus on the magnetic field-induced features, which are revealed by normalizing the spectrum at finite field by that at zero field. At a fixed magnetic field as low as 0.4 Tesla, we observed oscillations in the normalized spectrum that indicate the emergence of quantized LLs. These oscillations gradually evolve into several branches as we increase the magnetic field. These features correspond to inter-LL transitions, as expected from the schematic in Fig. 1c. The branch at the lowest energy is very broad in energy: we assign it to the transition 0 → 1 and -1 → 0. The breadth in energy of this transition is due to the lack of a vHS in the joint-density-of-states (jDOS) for this transition: 1. the zeroth LL lacks a DOS vHS due to its linear dispersion; 2. transitions between the almost parallel part of the $N$=0 and $N$=±1 LL bands at large $k_{//B}$ should have a vHS in jDOS, but they are forbidden by Pauli blocking (i.e. all states are either unoccupied or occupied in the region where $N$=0 and $N$=±1 are parallel). All other branches of magnetic field-induced features are sharp in energy due to vHSs (at the same $k_{//B}$) for both the initial and final states, corresponding to the inter-LL transitions. For now, we assume that these transitions are from the W2 Weyl Fermi surfaces; we later confirm this assumption by examining the scaling with field and the extracted Fermi velocity. We tentatively assign the dispersing features to pairs of LL indices, each corresponding to $\Delta|N|= \pm 1$, as shown in Fig. 1d.

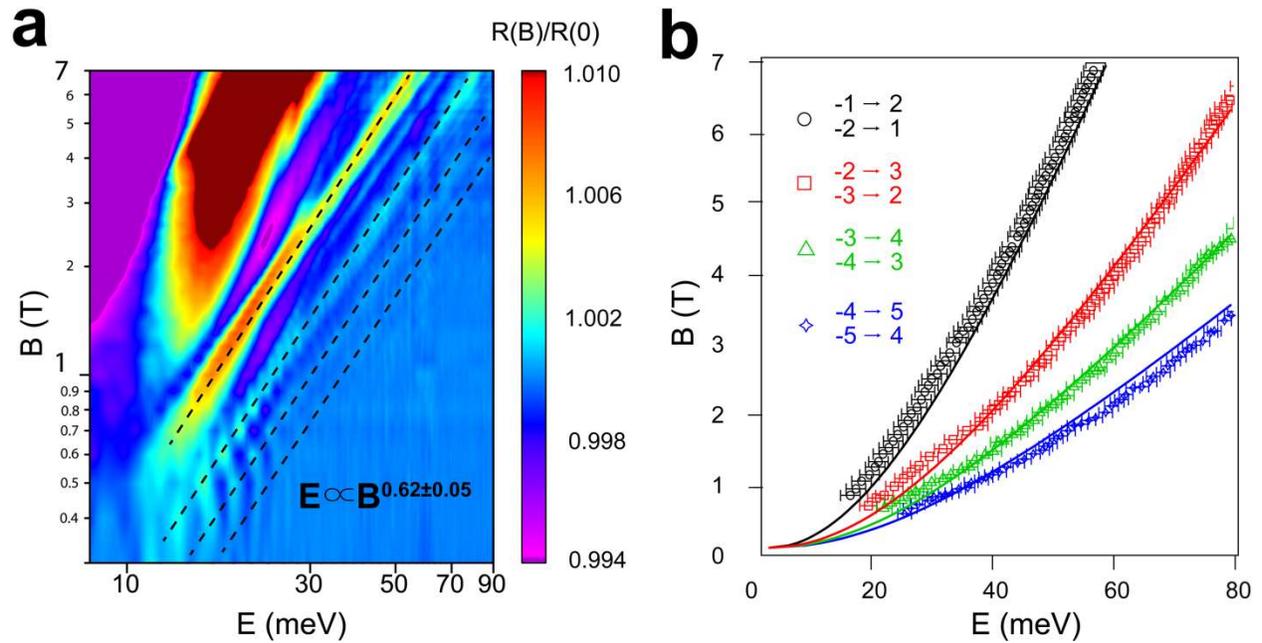

**Fig. 2 Scaling of inter-LL transition energies with magnetic field. a.** *2D color plot of the normalized reflectance spectrum in the range of B = 0.3 to 7 T. Both axes are plotted on a log scale. The color scale is the same as in Fig. 1d. The dominant inter-LL transitions can be described phenomenologically by a scaling of $E \sim B^{0.62 \pm 0.05}$, as indicated by dashed lines. The deviation of the scaling from $B^{0.5}$ - expected for a perfectly linearly-dispersing bands near the Weyl points – is expected due to the curvature of the band near the saddle-point that connects the two Weyl points (see SI for details).* **b.** *Comparison between the experimental*

inter-LL transition energies and those calculated from a model Hamiltonian that incorporates the saddle point (see SI for details). The four branches of features in Fig. 1d and 2a agree well with calculated dispersions plotted in black, red, green, and blue, from low to high energy. These transitions are attributed light polarization perpendicular to B.

Next, we examine the scaling relation of the inter-LL transition energies as a function of magnetic field. Fig. 2a shows the normalized reflectance on a log-log plot. The four major branches of features can be fit with a scaling of $E \propto B^{0.62\pm0.05}$. This is close to the $E \propto B^{0.5}$ scaling expected for the perfect linear dispersion of ideal Weyl cones, and distinct from the $E \propto B$ scaling expected for trivial, parabolic bands. This rules out the trivial hole pocket H1 as the source of the LL transitions.

To further clarify the origins of the inter-LL transitions, and to understand the precise origin of the $\propto B^{0.62\pm0.05}$ scaling, we use a two-band Hamiltonian that describes the electronic structure of Weyl points separated along $k_{//B}$ (11, 12):

$$H(k) = a(k_w^2 - k^2)\sigma_z + \hbar v_F(k_x\sigma_x + k_y\sigma_y), \tag{1}$$

where $\sigma_i$ is the $i^{th}$ Pauli matrix and $v_F$ is the Fermi velocity. The two Weyl nodes sit at $(0, 0, \pm k_w)$. This minimal Hamiltonian gives a global description of a pair of Weyl nodes with opposite chirality and preserves all of their topological properties (12). This Hamiltonian can be further reduced to

$$H(k) = -ak^2\sigma_z + \hbar v_F k \cdot \sigma \tag{2}$$

near each Weyl point by redefining $k \pm k_w \rightarrow k$, where $a$ characterizes the height of the saddle point between the two Weyl nodes (*see Supplementary Information figure S1*). Equation 2 can be quantized analytically in the presence of a magnetic field and from the resultant LL energies we reproduce dispersions of the experimentally observed transitions, as shown in Fig. 2b. The four major transitions traced by dashed lines in Fig. 1d and 2a agreed well with the calculated $\Delta|N|=\pm1$ transitions. They are excited by light polarization perpendicular to the $B$ field. Note that the correction to the $B^{0.5}$ scaling expected for perfectly linear bands emerges naturally in this model by incorporating curvature along the $k_w$ direction. At higher energy—above 100 meV—the dispersion deviates from $B^{0.62}$ scaling but the model that incorporates the saddle-point continues to describe the inter-LL transitions up to the highest measured magnetic field of 17.5 T (see Supplementary Information figure S5 for additional data). In addition, there are branches of weaker features in Fig. 1d and 2a whose energy dispersions are close to those calculated for $\Delta|N|=0$ transitions. It is likely that these branches of features are excited by light polarization parallel to the $B$ field, as our light source is largely unpolarized (see Supplementary Information Section 5 for further discussion) (18).

The Fermi velocity we extract by fitting the LL transition spectra is $2.2 \pm 0.1 \times 10^5$ m/s, which agrees well with the Fermi velocity at the W2 Weyl point extracted from previous quantum oscillation measurements and theoretical predictions *(8,9,20,21)*. The extracted Fermi velocity is one order of magnitude larger than what is expected for W1 Weyl point in this field orientation *(9)*. It is also more than a factor of two smaller than the Fermi velocity measured via quantum oscillations for the trivial hole pocket *(19)* which, in addition, should have LLs that disperse linearly or near-linearly in field. We therefore conclude that all major features observed in our experiment correspond to inter-LL transitions from the W2 Weyl pockets, and not from the W1 Weyl pockets or the trivial H1 pockets. We find that the curvature parameter in equation 2, $a = 0.5$ eV nm$^2$, provides the best fit to the data.

In addition to characterizing the band dispersion around the Weyl points, inter-LL transitions provide a direct way to measure the distance of the Fermi energy, $E_F$, from the Weyl points. The Fermi surface areas for the W1 and H2 pockets in TaAs are consistent across multiple groups *(19–22)*, suggesting that $E_F$ is determined by band alignment within the crystal and that impurity doping plays a minimal role in setting the chemical potential *(17)*. The position of $E_F$ directly determines the threshold magnetic field at which the quantum limit is reached—when all LL bands other than $N=0$ are either filled or empty, leaving only the two $N=0$ LL bands with opposite chirality partially filled. Fig. 3a illustrates how $E_F$ can be measured by the inter-LL transitions. We assume that the Weyl cone is electron-like so that the Fermi level is above the Weyl point, consistent with previous work *(8,9,20,21)*, although our results are insensitive to this assignment. At low magnetic fields, inter-LL transitions below Fermi level at $k_{//B} = \pm k_w$ are forbidden due to Pauli blocking—there can be no transitions from an occupied state to another occupied state. The corresponding branch of transitions will therefore not show up in the reflectance spectrum at low field. As $B$ increases to a critical value, the $N$th LL band edge is raised to the Fermi level, unblocking the $-(N-1) \rightarrow N$ transition. The corresponding branch of inter-LL transition features will then emerge in the reflectance spectrum as the field is increased beyond the critical value. We measure the threshold fields and energies of the inter-LL transitions to extract the distance from the Weyl node to $E_F$, as shown in Fig. 3a. Here we assume the band around the Weyl point has a linear dispersion for simplicity. Non-linear corrections to the dispersion used in Fig. 2c are smaller than the uncertainty and thus negligible (see Supplementary Information Section 3).

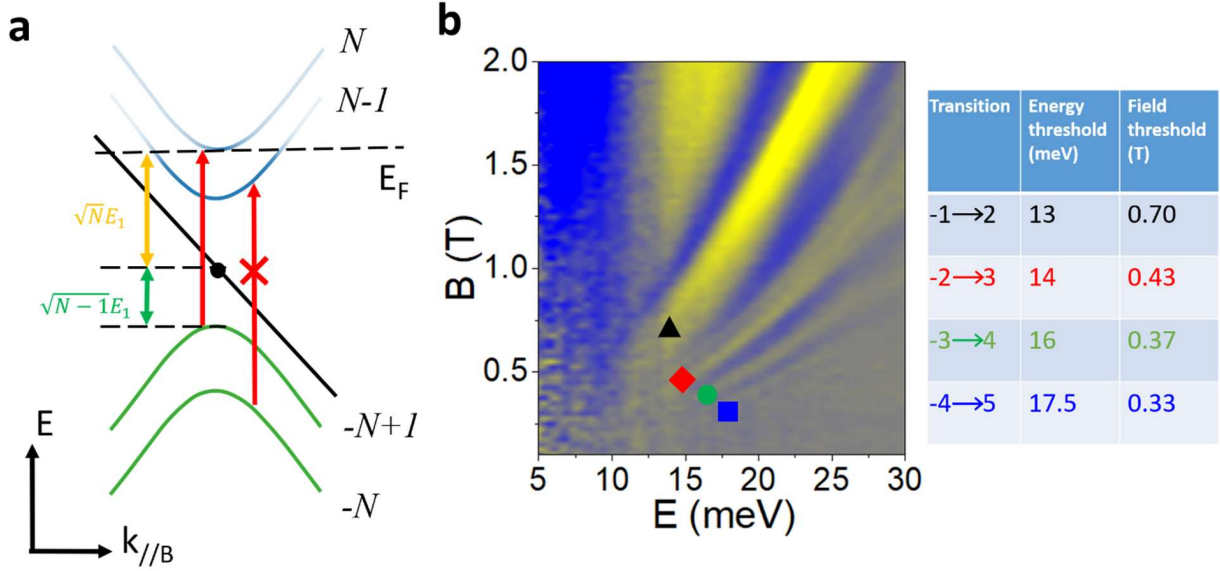

**Fig. 3 Extraction of Fermi energy and quantum limit of the Fermi surface at the W2 Weyl points. a.** *Illustration of the scheme for measuring the distance between the Weyl points and the Fermi energy, $E_F$. When B is increased so that the band edge of the Nth LL band is aligned with $E_F$, the inter-LL-band transition $-(N-1) \rightarrow N$ is no longer blocked. The corresponding branch appears in the reflectance spectrum at an energy $E = E_F(1 + \frac{\sqrt{N-1}}{\sqrt{N}})$. $E_1$ is the energy of the 1st LL band edge measured from the Weyl point (WP) at zero energy. The quantum limit is reached when $E_F = E_1$. **b.** 2D color plot of the normalized reflectance spectrum in the range of 0.1-2 T and 5-30 meV. The black triangle, red diamond, green dot, and purple square indicate the threshold fields of the four main branches. Their corresponding energies and magnetic fields are listed to the right of the plot.*

Figure 3b shows a 2D plot of inter-LL transitions in the range of $B$= 0.1-2 T. This data was taken in much finer steps than in Fig. 1d to reveal the details of the onset of the inter-LL transitions. Each major branch of features emerges beyond a threshold field and energy. Using the energies of these threshold points, we estimate the distance between $E_F$ and the Weyl points as 7.6, 7.7, 8.6 and 9.2 meV (from the -1→2, -2→3, -3→4, and -4→5 branches, respectively). We therefore concluded that $E_F$ is 8.3 ± 0.9 meV above the W2 Weyl points in TaAs. We further determine the magnetic field at which the quantum limit is reached to be $B_{QL} = B_N * \sqrt{N}$ (by definition, the quantum limit is reached when the 1st LL band edge emerges from below $E_F$.) Using the threshold magnetic field $B_N$ obtained from Fig. 3b, we extract the values of $B_{QL}$ as 0.99, 0.75, 0.74 and 0.74 T respectively. Thus, we conclude that the quantum limit is $B_{QL}$ = 0.8 ± 0.1T for Fermi surface around the W2 Weyl points.

A quantum limit of 0.8 T is significantly lower than has previously been reported for any other Weyl semimetal *(19,22–24)*. This includes prior studies of TaAs that compare quantum oscillations with band structure calculations, which suggested that the quantum limit of the W2 Fermi surface is between 5 and 8 Tesla, depending on the field orientation *(19,22)*. Our measurements suggest that previous identifications of the W2 pocket are incorrect. This is likely due to the difficulty in uniquely assigning calculated quantum oscillation frequencies to branches in the measured oscillation spectrum—a task made particularly difficult in TaAs because both the W1 and W2 Weyl pockets have nearly the same cross-sectional area when the magnetic field is applied along the *c*-axis.

To confirm that the other Fermi surfaces—the electron pockets centered around the W1 Weyl points and the hole pockets H1—have the same Fermi surface volume in our samples as reported previously, we measured quantum oscillations in the same sample used to perform the magneto-infrared optical measurements. We use pulse-echo ultrasound to measure quantum oscillations in the sound velocity. For field along the *c*-axis, we see a dominant oscillation frequency of approximately 7.5 tesla, as reported previously (identified as W1 in Ref. *(19)* and W2 in Ref. *(22)* —the same Fermi surface but with reversed nomenclature). With the field along the *a*-axis, we find a dominant oscillation frequency of approximately 1.4 telsa. Both the mass and quantum oscillation frequency of this 1.4 tesla pocket are consistent with the trivial hole pocket identified in Ref. *(19)* (see SI section 7 for details). There is a peak in the Fourier transform of the quantum oscillations near 0.8 tesla that may correspond to the W2 pocket, but the frequency is so low that it is difficult to distinguish from the background. This highlights the difficulty in using quantum oscillations to identify Weyl fermions: extremely low-frequency oscillations are lost in the background and the damping factor due to scattering makes them even more difficult to observe before they reach their quantum limit.

Having established the main features of the magneto-electrodynamics of the Weyl Fermions at the W2 Weyl points, we now describe several experimental observations that allow for more accurate descriptions of the Fermi surface at W2. Fig. 4a compares the inter-LL transition energies at three different in-plane angles $\theta$ over the same range of magnetic field. The redshift of the -1→2 & -2→1 branch of transitions shows a 20% relative change in energy as the field is rotated from $\theta = 45°$ to $\theta = 0°$. This shift indicates anisotropy in the *ab*-plane of momentum space for the W2 pockets. We propose an anisotropic shape for the W2 pockets as shown in Fig. 4b, consistent with the four-fold rotation and mirror reflection crystal symmetries of TaAs for this sample geometry. When the magnetic field is oriented 45° from the *a* or *b* axes, the cross-section perpendicular to *B* is smaller than when the field is pointing along the *a* or *b* axes. This explains the angle-dependence of inter-LL transition energies in Fig. 4a. While two-fold symmetry in the ab-plane is allowed for the W2 pockets based on their position in the Brillouin zone, two-fold anisotropy would result in a qualitatively different $\theta$–dependence: any two-fold anisotropy must be below our experimental resolution (see Supplementary Figure S2).

In addition to the anisotropy of the W2 Fermi surfaces, we observe fine features within the same branch of the inter-LL transitions, as shown in Fig. 4c. The peaks are split with an energy difference of approximately 8%, possibly indicating the breakdown of electron-hole symmetry and thus breaking the degeneracy between the –(N-1)→N and -N→(N-1) transitions.

Finally, to quantify the crystal quality, we provide an estimate of the scattering rate of the Weyl Fermions in the W2 pockets. As shown in Fig. 4d, three dispersions of the inter-LL transition peaks can be clearly resolved in the 2D color plot. By cutting a line at $B = 1.1$ T, we obtain an average peak separation of approximately 1.5 meV. Using the Rayleigh criteria, we conclude that the inhomogeneous broadening of inter-LL transition is less than 3 meV, corresponding to a scattering rate of 4.5 ps$^{-1}$ and a mean free path of 50 nm. This agrees well with the scattering rate measured via quantum oscillations in our samples grown previously (20). In addition, a high electron mobility ($\mu$ >30000 $cm^2/V \cdot s$) is extracted from the $\mu \cdot B \approx 1$ criteria for the emergence of the Landau level transitions at less than 0.3 T.

**Discussion**

We observed clear inter-LL transitions from the Fermi surface surrounding the W2 Weyl points of TaAs by utilizing high-quality TaAs crystals and the Voigt configuration for magneto-infrared reflection spectroscopy. The observed scaling laws of the transition energies confirm an electronic dispersion that is characteristic of Weyl fermions, and we extract the curvature parameter that leads to small deviations from linear dispersion due to the small separation between the two Weyl points. Surprisingly, we find that the magnetic field where the quantum limit is reached for this pocket is only 0.8 tesla – an order of magnitude smaller than previously reported *(19,22–24)*. To the best of our knowledge, this is the lowest quantum limit field reported for any Weyl semimetal to-date. This low quantum limit, and correspondingly low Fermi energy, makes TaAs favorable for further investigations of Weyl fermions. For example, one could incorporate a DC electric field in the ab-plane into our measurement scheme using thin, fabricated TaAs devices or films. This would pump charge from one Weyl point to its neighbor of opposite chirality, raising the chemical potential of one point with respect to the other and splitting the single onset at the quantum limit into two. Such charge-pumping is known as the chiral anomaly of Weyl Fermions *(18,25–28)*. Previous measurements of this effect relied on DC transport, which suffered from ambiguity due to other possible trivial effects (*29–31*). Our experiment suggests a more direct, spectroscopic measurement of this effect by directly probing the change in the chemical potential at each Weyl node induced by charge pumping. In addition, the distinct chirality and unique spin textures surrounding each Weyl point give rise to a new degree of freedom that is analogous to the valley degree of freedom in semiconductors. This has been studied through various techniques including photocurrent (*32*) and spin resolved ARPES (*33*). By introducing a magnetic field in a manner similar to what we have done here, the chiral LL states with energy separation in the THz range may offer a new scheme for optoelectronic and spintronic devices. Finally, we suggest that the energy-space separation offered by the LL spectroscopy we employ here is ideal for investigating the Weyl fermi surfaces of other materials where the fermiology is undetermined and likely complex, such as Mn$_3$Sn and Co$_2$MnGa.

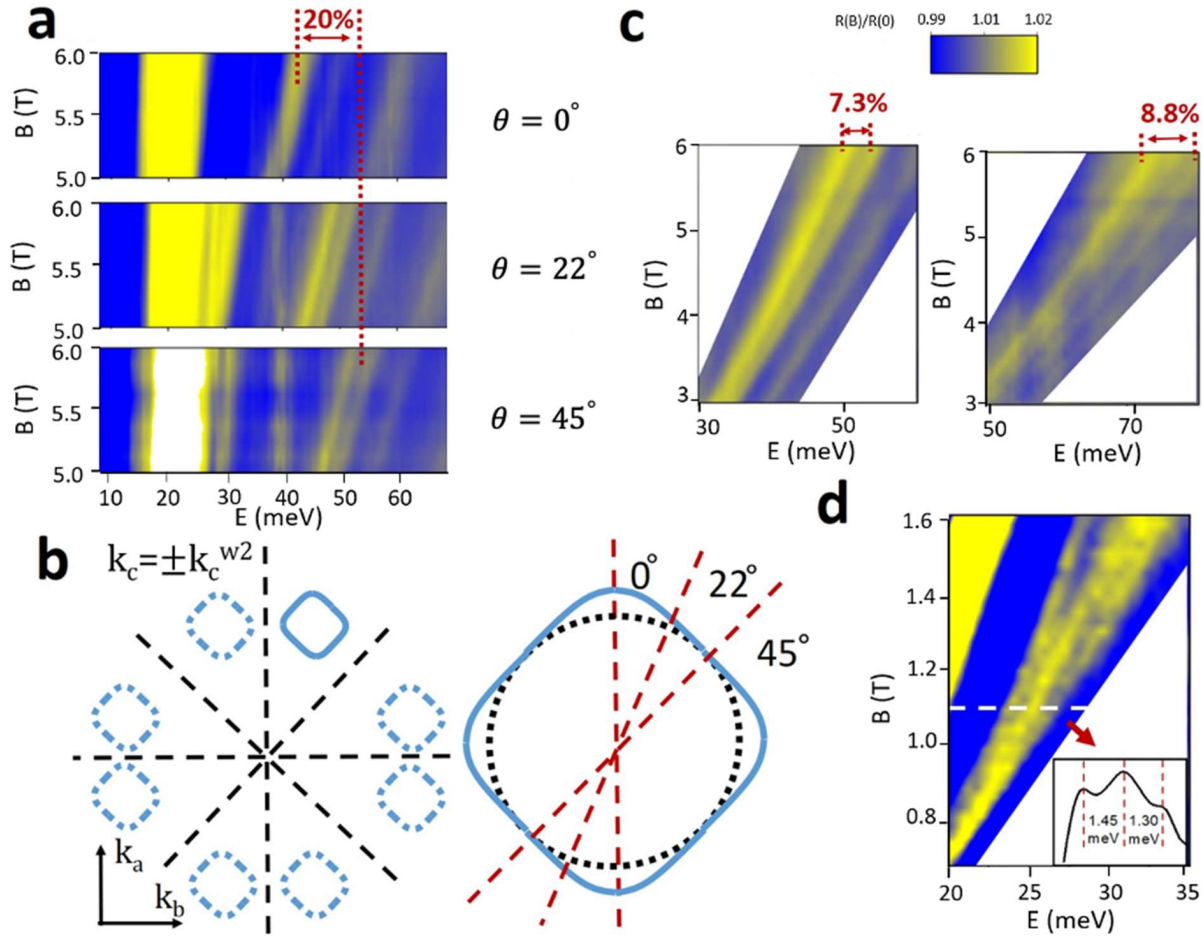

***Fig. 4 Characterization of the anisotropy and fine features of the W2 Weyl Fermi surface. a.*** *2D color plots of the normalized reflectivity in the range of 5-6 T for three different angles: $\theta = 0°$, 22° and 45°. The*

*color scale is shown in **c**, but shared by **a**, **c**, & **d**. The -1→2 and -2→1 branches redshift as the field is rotated from θ = 45° to θ = 0°, showing a relative change of 20%. **b**. Possible shape of W2 Weyl pockets in the ab-plane of the momentum space. W2 Weyl pockets lie in the $k_a$-$k_b$ plane at a constant $k_c=\pm k_c^{w2}$. Each pocket must have smaller (bigger) cross-section along the 45° (0°) direction to explain the angle-dependence of inter-LL transition energies. Due to the four-fold rotation symmetry and mirror reflection symmetries of the crystal lattice in the ab-plane, the square-shaped W2 pockets is the only possible shape to explain the data in **a** (a lower, two-fold symmetric pocket is allowed by symmetry but would produce features not seen in the data—any two-fold distortion must be below our resolution.). **c**. Fine splitting within the -1→2 & -2→1 branch and the -2→3 & -3→2 branch at θ = 45°, indicating possible electron-hole asymmetry. **d**. A linecut of Fig. 3b at B = 1.1 T. Three fine peaks, separated by ~1.5 meV, can be resolved. This observation implies a linewidth of inter-LL transitions on the order of 1.5 meV and small inhomogeneous broadening.*

**Acknowledgements:** B. J. R. and P. M. H. acknowledge support from the National Science Foundation under grant no. DMR-1752784. A portion of this work was performed at the National High Magnetic Field Laboratory, which is supported by the National Science Foundation Cooperative Agreement No. DMR-1644779 and the State of Florida. Work at Los Alamos was performed under the auspices of the U.S. Department of Energy, Office of Basic Energy Sciences, Division of Materials Science and Engineering. **Competing interests:** All authors declare that they have no competing interests. **Author contributions:** B. J. R and L. J. conceived the work. Samples were grown by E. D. B. and F. R. B. J. R. polished and oriented the samples. Infrared spectroscopy measurements were performed by L. J., Z. L., M. O., S. M., and D. S. P. M. H performed the ultrasound measurements. Z. L., L. J., and B. J. R. wrote the manuscript with contributions from all co-authors. **Data and Materials Availability:** All data needed to evaluate the conclusions in the paper are present in the paper and/or the Supplementary Materials.

## Methods

**Sample growth and orientation**.

Millimeter-sized single crystals of TaAs were grown by means of an iodine-vapor transport technique with 0.05 g/cm$^3$ I$_2$. First, polycrystalline TaAs was prepared by heating stoichiometric amounts of Ta and As in an evacuated silica ampoule at 973 K for three days. Subsequently, the powder was loaded in a horizontal tube furnace in which the temperature of the hot zone was kept at 1123 K and that of the cold zone was ~1023 K. The composition and structure of the TaAs single crystals was verified by Laue diffraction and energy dispersive x-ray spectroscopy (EDS). No I$_2$ doping was detected.

Crystals were approximately aligned by their morphological structure, and the alignment was refined to better than 1° using Laue diffraction. Flat, ab-plane surfaces (perpendicular to the tetragonal c-axis) were polished using 1 micron diamond lapping film and finished with 30 nm lapping film.

**Magneto-infrared reflectance measurements**.

Broadband magneto-infrared measurements were performed using a Bruker 80v Fourier-transform infrared spectrometer at liquid helium temperature with a superconducting magnet up to 17.5 T. The unpolarized broadband light from a globar was delivered to the sample through evacuated light pipes and focused on the sample in Voigt geometry. The sample was sitting at the field center in a helium exchange gas environment. The reflected light was collected and delivered to a Si bolometer placed away from the magnetic field center.

1.  B. Yan, C. Felser, Topological Materials: Weyl Semimetals. *Annu. Rev. Condens. Matter Phys.* **8**, 337–354 (2017).

2.  A. Bansil, H. Lin, T. Das, Colloquium: Topological band theory. *Rev. Mod. Phys.* **88**, 21004 (2016).

3.  N. P. Armitage, E. J. Mele, A. Vishwanath, Weyl and Dirac semimetals in three-dimensional solids. *Rev. Mod. Phys.* (2018), doi:10.1103/RevModPhys.90.015001.

4.  S.-Y. Xu, I. Belopolski, N. Alidoust, M. Neupane, G. Bian, C. Zhang, R. Sankar, G. Chang, Z. Yuan, C.-C. Lee, S.-M. Huang, H. Zheng, J. Ma, D. S. Sanchez, B. Wang, A. Bansil, F. Chou, P. P. Shibayev, H. Lin, S. Jia, M. Z. Hasan, Discovery of a Weyl fermion semimetal and topological Fermi arcs. *Science (80-. )*. **349**, 613 (2015).

5.  B. Q. Lv, H. M. Weng, B. B. Fu, X. P. Wang, H. Miao, J. Ma, P. Richard, X. C. Huang, L. X. Zhao, G. F. Chen, Z. Fang, X. Dai, T. Qian, H. Ding, Experimental Discovery of Weyl Semimetal TaAs. *Phys. Rev. X*. **5**, 31013 (2015).

6.  H. Weyl, Elektron und Gravitation. I. *Zeitschrift für Phys.* **56**, 330–352 (1929).

7.  Y. Li, F. D. M. Haldane, Topological Nodal Cooper Pairing in Doped Weyl Metals (2018), doi:10.1103/PhysRevLett.120.067003.

8.  D. Grassano, O. Pulci, A. Mosca Conte, F. Bechstedt, Validity of Weyl fermion picture for

transition metals monopnictides TaAs, TaP, NbAs, and NbP from ab initio studies. *Sci. Rep.* **8**, 3534 (2018).

9. C.-C. Lee, S.-Y. Xu, S.-M. Huang, D. S. Sanchez, I. Belopolski, G. Chang, G. Bian, N. Alidoust, H. Zheng, M. Neupane, B. Wang, A. Bansil, M. Z. Hasan, H. Lin, Fermi surface interconnectivity and topology in Weyl fermion semimetals TaAs, TaP, NbAs, and NbP. *Phys. Rev. B*. **92**, 235104 (2015).

10. S. Kimura, H. Yokoyama, H. Watanabe, J. Sichelschmidt, V. Sü\ss, M. Schmidt, C. Felser, Optical signature of Weyl electronic structures in tantalum pnictides $\mathrm{Ta}Pn$ ($Pn=$ P, As). *Phys. Rev. B*. **96**, 75119 (2017).

11. S.-B. Zhang, H.-Z. Lu, S.-Q. Shen, Linear magnetoconductivity in an intrinsic topological Weyl semimetal. *New J. Phys.* **18**, 53039 (2016).

12. H.-Z. Lu, S.-B. Zhang, S.-Q. Shen, High-field magnetoconductivity of topological semimetals with short-range potential. *Phys. Rev. B*. **92**, 45203 (2015).

13. M. Orlita, D. M. Basko, M. S. Zholudev, F. Teppe, W. Knap, V. I. Gavrilenko, N. N. Mikhailov, S. A. Dvoretskii, P. Neugebauer, C. Faugeras, A.-L. Barra, G. Martinez, M. Potemski, Observation of three-dimensional massless Kane fermions in a zinc-blende crystal. *Nat. Phys.* **10**, 233–238 (2014).

14. Y. Jiang, Z. Dun, S. Moon, H. Zhou, M. Koshino, D. Smirnov, Z. Jiang, Landau Quantization in Coupled Weyl Points: A Case Study of Semimetal NbP. *Nano Lett.* **18**, 7726–7731 (2018).

15. X. Yuan, Z. Yan, C. Song, M. Zhang, Z. Li, C. Zhang, Y. Liu, W. Wang, M. Zhao, Z. Lin, T. Xie, J. Ludwig, Y. Jiang, X. Zhang, C. Shang, Z. Ye, J. Wang, F. Chen, Z. Xia, D. Smirnov, X. Chen, Z. Wang, H. Yan, F. Xiu, Chiral Landau levels in Weyl semimetal NbAs with multiple topological carriers. *Nat. Commun.* **9**, 1854 (2018).

16. S. Kimura, Y. Yokoyama, Y. Nakajima, H. Watanabe, J. Sichelschmidt, V. Süß, M. Schmidt, C. Felser, in *Proceedings of the International Conference on Strongly Correlated Electron Systems (SCES2019)* (Journal of the Physical Society of Japan, 2020), vol. 30 of *JPS Conference Proceedings*.

17. S. Polatkan, M. O. Goerbig, J. Wyzula, R. Kemmler, L. Z. Maulana, B. A. Piot, I. Crassee, A. Akrap, C. Shekhar, C. Felser, M. Dressel, A. V Pronin, M. Orlita, Magneto-Optics of a Weyl Semimetal beyond the Conical Band Approximation: Case Study of TaP. *Phys. Rev. Lett.* **124**,

176402 (2020).

18. A. L. Levy, A. B. Sushkov, F. Liu, B. Shen, N. Ni, H. D. Drew, G. S. Jenkins, Optical evidence of the chiral magnetic anomaly in the Weyl semimetal TaAs. *Phys. Rev. B*. **101**, 125102 (2020).

19. F. Arnold, M. Naumann, S.-C. Wu, Y. Sun, M. Schmidt, H. Borrmann, C. Felser, B. Yan, E. Hassinger, Chiral Weyl Pockets and Fermi Surface Topology of the Weyl Semimetal TaAs. *Phys. Rev. Lett.* **117**, 146401 (2016).

20. B. J. Ramshaw, K. A. Modic, A. Shekhter, Y. Zhang, E.-A. Kim, P. J. W. Moll, M. D. Bachmann, M. K. Chan, J. B. Betts, F. Balakirev, A. Migliori, N. J. Ghimire, E. D. Bauer, F. Ronning, R. D. McDonald, Quantum limit transport and destruction of the Weyl nodes in TaAs. *Nat. Commun.* **9**, 2217 (2018).

21. X. Huang, L. Zhao, Y. Long, P. Wang, D. Chen, Z. Yang, H. Liang, M. Xue, H. Weng, Z. Fang, X. Dai, G. Chen, Observation of the Chiral-Anomaly-Induced Negative Magnetoresistance in 3D Weyl Semimetal TaAs. *Phys. Rev. X*. **5**, 31023 (2015).

22. C.-L. Zhang, B. Tong, Z. Yuan, Z. Lin, J. Wang, J. Zhang, C.-Y. Xi, Z. Wang, S. Jia, C. Zhang, Signature of chiral fermion instability in the Weyl semimetal TaAs above the quantum limit. *Phys. Rev. B*. **94**, 205120 (2016).

23. P. J. W. Moll, A. C. Potter, N. L. Nair, B. J. Ramshaw, K. A. Modic, S. Riggs, B. Zeng, N. J. Ghimire, E. D. Bauer, R. Kealhofer, F. Ronning, J. G. Analytis, Magnetic torque anomaly in the quantum limit of Weyl semimetals. *Nat. Commun.* **7**, 12492 (2016).

24. K. A. Modic, T. Meng, F. Ronning, E. D. Bauer, P. J. W. Moll, B. J. Ramshaw, Thermodynamic Signatures of Weyl Fermions in NbP. *Sci. Rep.* **9**, 2095 (2019).

25. M. M. Jadidi, M. Kargarian, M. Mittendorff, Y. Aytac, B. Shen, J. C. König-Otto, S. Winnerl, N. Ni, A. L. Gaeta, T. E. Murphy, H. D. Drew, Nonlinear optical control of chiral charge pumping in a topological Weyl semimetal. *Phys. Rev. B*. **102**, 245123 (2020).

26. H. B. Nielsen, M. Ninomiya, The Adler-Bell-Jackiw anomaly and Weyl fermions in a crystal. *Phys. Lett. B*. **130**, 389–396 (1983).

27. V. Aji, Adler-Bell-Jackiw anomaly in Weyl semimetals: Application to pyrochlore iridates. *Phys. Rev. B*. **85**, 241101 (2012).

28. A. A. Burkov, Chiral anomaly and transport in Weyl metals. *J. Phys. Condens. Matter*. **27**, 113201

(2015).

29. D. T. Son, B. Z. Spivak, Chiral anomaly and classical negative magnetoresistance of Weyl metals. *Phys. Rev. B*. **88**, 104412 (2013).

30. F. Arnold, C. Shekhar, S.-C. Wu, Y. Sun, R. D. dos Reis, N. Kumar, M. Naumann, M. O. Ajeesh, M. Schmidt, A. G. Grushin, J. H. Bardarson, M. Baenitz, D. Sokolov, H. Borrmann, M. Nicklas, C. Felser, E. Hassinger, B. Yan, Negative magnetoresistance without well-defined chirality in the Weyl semimetal TaP. *Nat. Commun.* **7**, 11615 (2016).

31. T. Schumann, M. Goyal, D. A. Kealhofer, S. Stemmer, Negative magnetoresistance due to conductivity fluctuations in films of the topological semimetal $Cd_3As_2$. *Phys. Rev. B*. **95**, 241113 (2017).

32. Q. Ma, S.-Y. Xu, C.-K. Chan, C.-L. Zhang, G. Chang, Y. Lin, W. Xie, T. Palacios, H. Lin, S. Jia, P. A. Lee, P. Jarillo-Herrero, N. Gedik, Direct optical detection of Weyl fermion chirality in a topological semimetal. *Nat. Phys.* **13**, 842–847 (2017).

33. S.-Y. Xu, I. Belopolski, D. S. Sanchez, M. Neupane, G. Chang, K. Yaji, Z. Yuan, C. Zhang, K. Kuroda, G. Bian, C. Guo, H. Lu, T.-R. Chang, N. Alidoust, H. Zheng, C.-C. Lee, S.-M. Huang, C.-H. Hsu, H.-T. Jeng, A. Bansil, T. Neupert, F. Komori, T. Kondo, S. Shin, H. Lin, S. Jia, M. Z. Hasan, Spin Polarization and Texture of the Fermi Arcs in the Weyl Fermion Semimetal TaAs. *Phys. Rev. Lett.* **116**, 96801 (2016).

34. Palik, E. D. & Furdyna, J. K. Infrared and microwave magnetoplasma effects in semiconductors. *Reports Prog. Phys.* **33**, 1193–1322 (1970).

35. Das Sarma, S. & Hwang, E. H. Collective Modes of the Massless Dirac Plasma. *Phys. Rev. Lett.* **102**, 206412 (2009).

36. Hofmann, J., Barnes, E. & Das Sarma, S. Interacting Dirac liquid in three-dimensional semimetals. *Phys. Rev. B* **92**, 45104 (2015).

37. Hofmann, J. & Das Sarma, S. Plasmon signature in Dirac-Weyl liquids. *Phys. Rev. B* **91**, 241108 (2015).

38. Hofmann, J. Quantum oscillations in Dirac magnetoplasmons. *Phys. Rev. B* **10**, 1–10 (2019).

39. Shao, J. M. & Yang, G. W. Magneto-optical conductivity of Weyl semimetals with quadratic term in momentum. *AIP Adv.* **6**, (2016).

# Supplementary Information: Weyl Fermion Magneto-Electrodynamics and Ultra-low Field Quantum Limit in TaAs

## 1. Low energy model and inter-LL transitions.

A minimal two-band model that describe the Weyl nodes can be written as *(12)*

$$H(k) = a(k_w^2 - k^2)\sigma_z + \hbar v_F(k_x\sigma_x + k_y\sigma_y), \tag{1}$$

where σ are the Pauli matrices, k is the wavevector, $v_F$ is the Fermi velocity, and *a* is a fitting parameter. This two band Hamiltonian gives a global description of the topological properties of a pair of Weyl nodes with opposite chirality *(12)*. The dispersion relations of these two energy bands are:

$$E(k)_\pm = \pm\sqrt{(a(k_w^2 - k^2))^2 + \hbar^2 v_F^2(k_x^2 + k_y^2)}. \tag{2}$$

At $k_x, k_y=0$, the two bands intersect at $(0, 0, \pm k_w)$, which are the Weyl points. If we apply a magnetic field in *z*-direction (along the direction of Weyl point separation), the electronic bands are quantized into a set of Landau levels dispersing along $k_z$. The dispersion relation is

$$E(k_z)^0 = \frac{a}{l_B^2} - ak_w^2 + ak_z^2, \quad N=0 \tag{3}$$

$$E(k_z)^N = \frac{a}{l_B^2} \pm \sqrt{(ak_w^2 - ak_z^2 - \frac{2a}{l_B^2}N)^2 + \frac{2\hbar^2 v_F^2}{l_B^2}N}, \quad N \geq 1, \tag{4}$$

where N is the Landau index and $l_B \equiv \sqrt{\hbar/|eB|}$ is the magnetic length. At $k_z = \pm k_w$, the dispersion relation for N≥1 reduced to

$$E(\pm k_w)^N = \frac{a}{l_B^2} \pm \sqrt{\frac{4a^2}{l_B^4}N^2 + \frac{2\hbar^2 v_F^2}{l_B^2}N}. \tag{5}$$

The corresponding inter-LL transition energy from –(N-1) to N is:

$$E(\pm k_w)^{-(N-1) \to N} = \sqrt{\frac{4a^2}{l_B^4}N^2 + \frac{2\hbar^2 v_F^2}{l_B^2}N} + \sqrt{\frac{4a^2}{l_B^4}(N-1)^2 + \frac{2\hbar^2 v_F^2}{l_B^2}(N-1)}. \tag{6}$$

Since the Hamiltonian is still particle-hole symmetric, the inter-LL transition from -N to (N-1) will be the same energy. This analytical solution only exists when magnetic field is applied along the Weyl node separation direction. For a magnetic field point to an arbitrary direction in ab plane, one can utilize an effective Hamiltonian reduced from the two-band model near each Weyl points. The Hamiltonian describing a single Weyl cone near $k_z=\pm k_w$ is:

$$H(k) = -ak^2\sigma_z + \hbar v_F k \cdot \sigma, \tag{7}$$

As a result, the LL energy shares the same form as the two-band model at $k_z=k_w$:

$$E^0 = -\frac{a}{l_B^2}, \quad N=0 \tag{8}$$

$$E^N = -\frac{a}{l_B^2} \pm \sqrt{(\frac{2a}{l_B^2}N)^2 + \frac{2\hbar^2 v_F^2}{l_B^2}N}, \quad N \geq 1 \tag{9}$$

as well as the inter-LL transition energy:

$$E^{-(N-1)\to N} = \sqrt{\frac{4a^2}{l_B^4}N^2 + \frac{2\hbar^2 v_F^2}{l_B^2}N} + \sqrt{\frac{4a^2}{l_B^4}(N-1)^2 + \frac{2\hbar^2 v_F^2}{l_B^2}(N-1)}. \tag{10}$$

Since this single Weyl point Hamiltonian is symmetric along $k_x$, $k_y$, $k_z$, the actual form of the inter-LL transition does not depend on the orientation of the magnetic field direction. In particular, the inter-band transitions we are interested in are at/near the Weyl points, so eq. S10 will still be valid for fitting the data sets with magnetic field 22 deg and 45 deg to the a-axis. In addition, if we consider the particle-hole asymmetry, the Hamiltonian can be further modified as:

$$H(k) = \frac{\hbar^2 k^2}{2m^*} - ak^2\sigma_z + \hbar v_F k \cdot \sigma, \tag{11}$$

which will give a small but finite energy splitting between the –(N-1)→N and -N→(N-1) transitions.

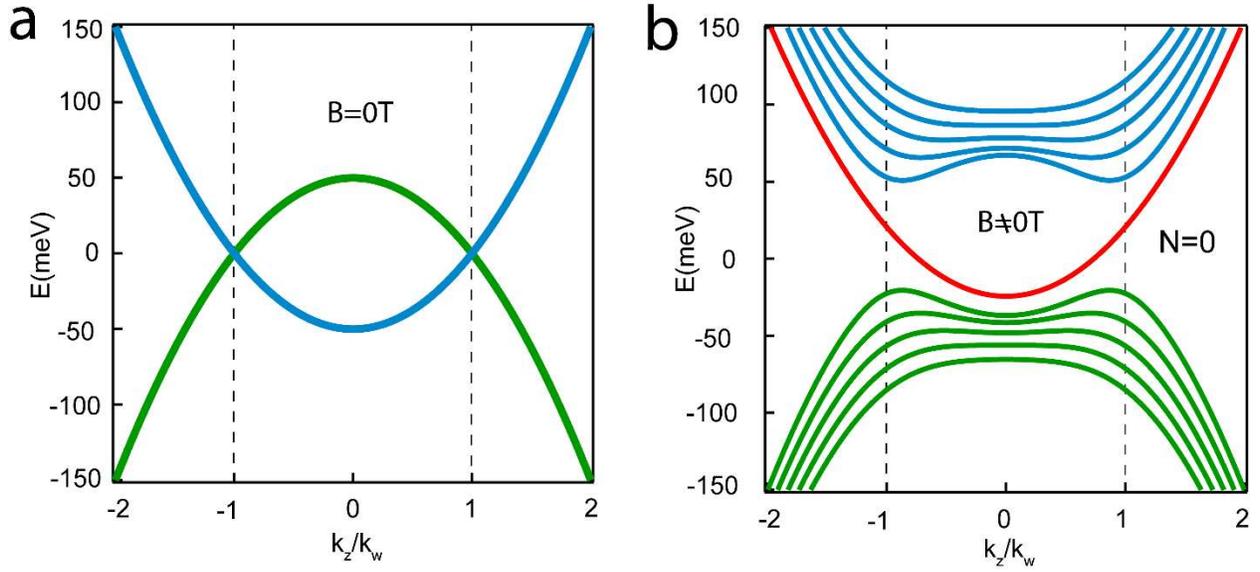

***Fig. S1 Low energy dispersion along $k_z$ with the two-band model. a.*** *Dispersion relation at zero magnetic field, Weyl nodes are sitting at $k_z=\pm k_w$, labeled with dashed lines.* ***b.*** *Landau levels at a finite magnetic field along z direction of both Weyl nodes. The N=0 Landau band is labeled in red.*

## 2. Anisotropy of the W2 pockets

Assuming that the pockets have two-fold rotation symmetry, like in an ellipse, a magnetic field pointing along either the 45° or 0° directions will result in two different cross-sections for the pockets. If the ellipse is elongated along the a/b axis, at $\theta = 45°$, all W2 pockets will have the same cross-section ($c_{45}$)—resulting in one inter LL transition energy for one branch. At $\theta = 0°$, half of the W2 pockets will have a smaller (bigger) cross-section than that at $\theta = 45°$, resulting in two different inter LL transition energies in the same branch that sandwich the transition energy at theta=45° (Fig. S2b). However, we did not observe such structures in Fig. 4a: instead, we see only a continuous redshift of the lines as we rotate from $\theta = 45°$ to $\theta = 0°$. Note that this is distinct from the small splitting we show in Figure 4c that may arise due to particle-hole asymmetry. Other possible Fermi surface shapes with two-fold rotation symmetry will give similar results and also do not agree with the data: in general, two-fold Fermi surfaces will always produce a feature that red-shifts when rotating away from $\theta = 0°$, which we do not observe.

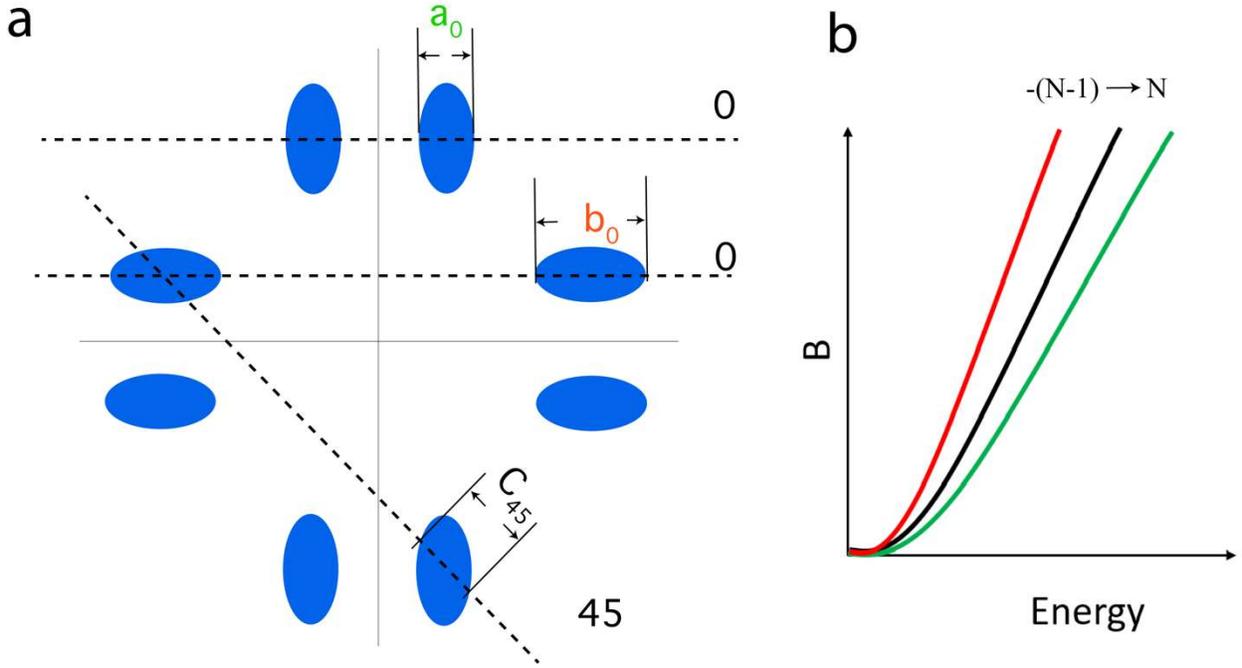

*Fig. S2 Schematic of anisotropic W2 pockets. **a**. The W2 Fermi surface, with each pocket having only C2 symmetry. $a_0$ and $b_0$ indicates the cross-section when magnetic field is aligned with a/b axis. $c_{45}$ is the cross-section when magnetic field is 45 deg to the a/b axis. **b**. Schematic of the inter LL transitions for magnetic field aligned with a/b axis (green and red, respectively) and 45 deg to either axis (black). The LL transition for the 45 deg case is sandwiched between the two other transitions, which is not what we observe in Figure 4a.*

### 3. Fermi energy correction induced by nonlinear dispersion

By considering the contribution from the non-linear dispersion, the Fermi energy can be estimated from: $E = E_F(1 + \frac{\sqrt{N-1+0.036(N-1)^2}}{\sqrt{N+0.036N^2}})$. Taking the -1→2 transition as an example gives a 0.05 meV correction for the Fermi energy which is smaller than the measurement uncertainty.

### 4. Estimation of plasmon-induced energy shift

In Voigt geometry, the transition frequencies will be shifted due to coupling with the plasmon and the actual resonance will occur at $\omega = \sqrt{\omega_0^2 + \omega_p^2}$ instead of $\omega_0$, where $\hbar\omega_0$ is the transition energy without coupling to the plasmon and $\omega_p$ is the plasmon frequency *(34)*. In a 3-dimenonal Weyl system, where the energy dispersion is linear, the plasmon frequency at zero temperature and finite density is proportional to the Fermi energy and $n^{1/3}$, where n is the carrier density *(35)*. The form of the plasmon frequency can be written as $\omega_p = \sqrt{\frac{2g\alpha}{3\pi\kappa_0}}E_F$, where $\alpha = \frac{k_e e^2}{\hbar v_F k}$ is a dimensionless constant (the ratio between coulomb and kinetic energy), $g$ is the degeneracy of the Weyl nodes, and $E_F$ is Fermi energy. The parameter $\kappa_0 = 1 + \frac{g\alpha}{3\pi}ln\frac{\Lambda_L}{2k_F}$, where $\Lambda_L = k_c e^{\frac{g\alpha}{3\pi}}$, is known as the Landau pole *(36)*, and $k_c$ is the cut-off corresponding to the momentum scale above which the dispersion deviates from linear. For Weyl semimetals like TaAs, with Weyl pairs in close proximity in the Brillouin zone, the Landau pole could be very small *(37)*. Therefore, the estimated energy of the plasmon is $\hbar\omega_p \approx 1.5$meV, which is much smaller than the transition energies we observed (>10 meV). Since the plasmon induced resonance frequency shift is less than few percent *(38)*, which is within the error bar of the data fitting in Figure 2.b, the effects induced by plasmon can be neglected.

## 5. Inter-LL transition assignment and fine features

Following the derivation of the optical selection rules *(18,39)*, the inter-LL transitions with $\Delta|N| = \pm 1$ or $\Delta|N| = 0$ are associated to light polarization perpendicular or parallel to the *B* field. In the experimental data, the main branches (Fig. 1d and Fig. 2a) are attributed to the transitions between $\Delta|N| = \pm 1$ LLs at the W2 Weyl points. Using the Hamiltonian described above, we can reproduce the dispersions with Fermi velocity $2.2\times10^5$ m/s for all branches. The four major transitions traced by dashed lines in Fig. 1d and 2a agreed well with the calculated $\Delta|N|= \pm 1$ transitions. Other than these strong features, a set of branches with weaker intensity are likely to be the $\Delta|N| = 0$ transitions, which are excited by light polarization parallel to the B field. By using the same fitting parameters, we can calculate the transition energies for the $\Delta|N| = 0$ transitions, which agree with the experimental data very well (Fig. S3a). If we compare the energies of the strong features with the calculated $\Delta|N| = 0$ transitions, they will not agree well even after optimizing the fitting parameters (Fig. S3b). Similarly, calculated $\Delta|N| = \pm 1$ transition energies do not agree with the weaker branches (Fig. S3c). As a result, we assign the stronger branches and weaker branches to the to $\Delta|N| = \pm 1$ and $\Delta|N| = 0$ inter-LL transitions, respectively.

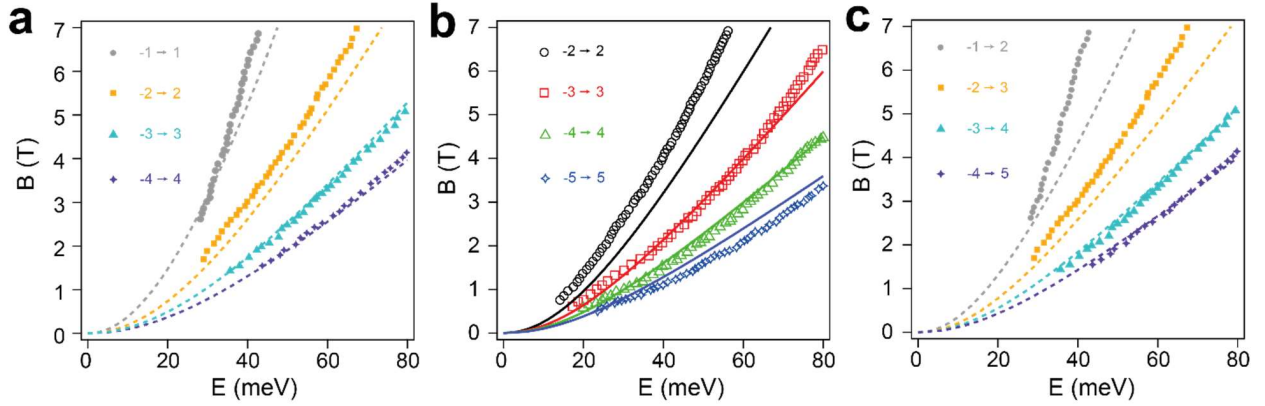

**Fig. S3 Fitting of the stronger and weaker features with different assignment of the inter-LL transitions.** *a. Comparison between the energies of weaker features extracted from Fig. 1d and calculated inter-LL transition energies of $\Delta|N| = 0$ with the same fitting parameters as used in the main text. The four branches of the weaker features agree well with the calculated energies plotted in grey, orange, cyan and purple dashed lines. b. Assigning the main features shown in Fig.1d to $\Delta|N| = 0$ transitions and comparing with the calculated inter-LL transition energies. The solid lines in black, red, green and blue are the best fit to the data with a Fermi velocity of $1.9 \times 10^5$ m/s. c. Fitting the weaker features to $\Delta|N| = \pm 1$ transitions with the same parameter as used in Fig. S3b. From these fits it is clear that the assignments of the strong features to $\Delta|N| = 0$ transitions and the weak features to $\Delta|N| = \pm 1$, does not produce as good of fits as the opposite assignment (which is the assignment used in the main text).*

Ideally, with perfectly unpolarized incidence light, the intensity of both $\Delta|N| = \pm 1$ and $\Delta|N| = 0$ transitions should be comparable following the derivation of the matrix elements *(18,39)*. However, in the experimental data, the features correspond to $\Delta|N| = 0$ transitions are weaker than those from $\Delta|N| = \pm 1$. A possible reason could be that although the incident light is largely unpolarized, the experimental apparatus could prefer one polarization more than the other due to the geometry of components and the sample. In our case, we suspect that light with polarization perpendicular to $B$ is stronger than the light with polarization parallel to $B$, which results in the intensity difference of the two set of features.

## 6. Normalized reflection spectra

Fig. S4 shows part of normalized reflection spectra that were used to generate Fig. 1d in the main text.

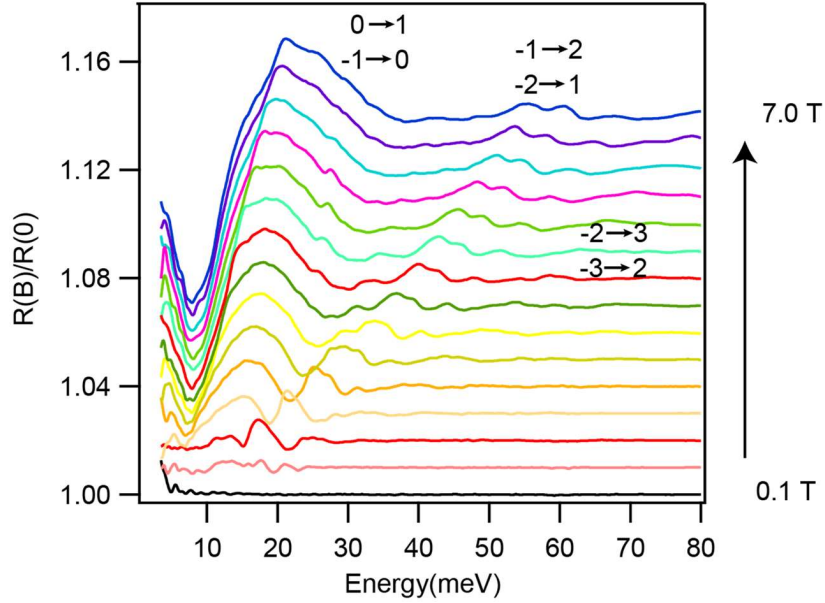

***Fig. S4 Reflection spectra normalized to zero field, R(B)/R(0), from 0.1 T to 7 T.*** *The labels indicate the corresponding inter-LL transitions. The spectra are vertically shifted for clarity.*

## 7. Estimation of the Energy scale of W2 Weyl cone

To estimate the energy scale of the W2 Weyl cone, we compared the fitting results based on the phenomenological model described in SI section 1, which includes deviations from the linear dispersion and incorporates the saddle points and the $B^{0.62}$ scaling. The data, including transitions at a higher energy at up to 17.5T, is shown in Figure. S5. The solid lines are fits that result from the phenomenological model with the same set of parameters mentioned in the main text. The dashed lines are calculated energy dispersion based on $B^{0.62}$ scaling. As we can see, below 80-90 meV, the data follows very well with both $B^{0.62}$ scaling and the phenomenological model. However, above 100meV, the magnetic field dependence of the transition energy deviates from $B^{0.62}$, while the full model continues to capture the dispersion relatively well. Such deviation from the simple scaling could be an indication that the energy scale is approaching the saddle point. In addition, we can calculate the energy scale of the saddle point from the phenomenological model with the extracted fitting parameters (a=0.5 eV nm$^2$). If we take $k_w$ = 0.5 nm$^{-1}$ from Ref. *(4)*, the energy of the saddle point is roughly 120 meV. Consequently, the energy scale of about 120 meV agrees with the theoretical prediction of W2 in Ref. *(8)*.

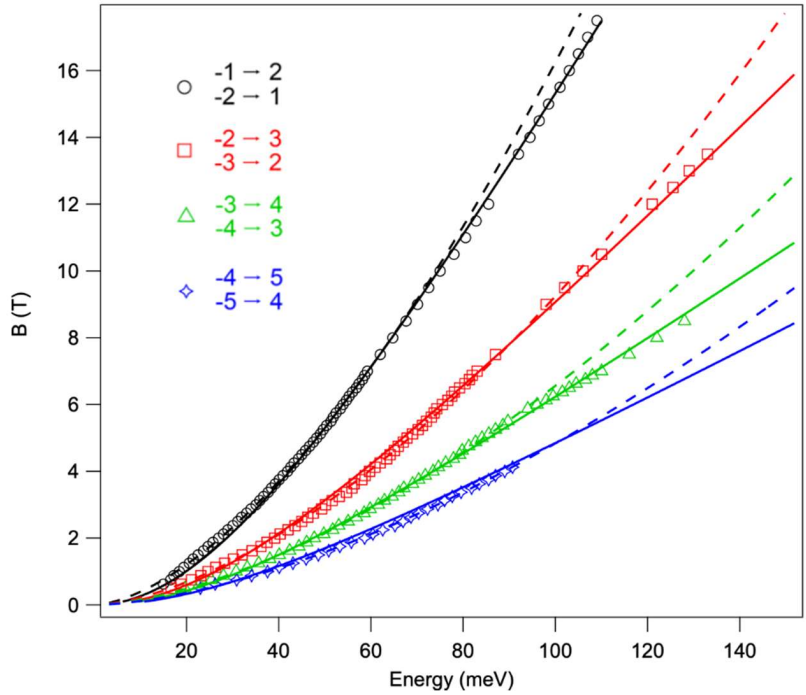

***Fig. S5 The Δ|N| = ±1 transitions up to 17.5 T.*** *The markers are LL-transitions extracted from the normalized reflection spectra. The dashed lines represent the dispersion based on the $B^{0.62}$ scaling. The solid lines are fits to the model described in section 1. Above 100meV, the magnetic field dependence of the transition energy deviates from $B^{0.62}$ but continues to fit the phenomenological model.*

## 8. Pulse echo ultrasound

A longitudinal thin-film ZnO transducer was reactively sputtered onto the flat a-b plane surface shown in Figure 1b of the main text. The relative change in the c-axis sound velocity was measured as a function of magnetic field with a digital pulse-echo technique described in Ref. *(20)*.

Fig. S6a shows quantum oscillations in the c-axis sound velocity for magnetic field applied along the [100] direction, corresponding to $\theta = 0°$. The oscillations are highly asymmetric around the background – either there is a very low frequency oscillation that is hard to resolve because of the dominant $F = 1.4$ T oscillation, or there is a non-trivial background due to movement of the chemical potential with field. Fig. S6b shows the Fourier transform of these data. The dominant peak at $F = 1.4$ T corresponds to the oscillations most visible by eye in panel a. The peak near $F = 0.8$ T is hard to resolve separately from the background at lower frequency, but is likely the source of asymmetry seen in panel a. This peak corresponds well with the quantum limit measured at the W2 Weyl points using magneto-infrared spectroscopy, but it is impossible to say anything conclusive about the Fermi surface using the quantum oscillation data alone because the very low frequency is difficult to distinguish from the background.

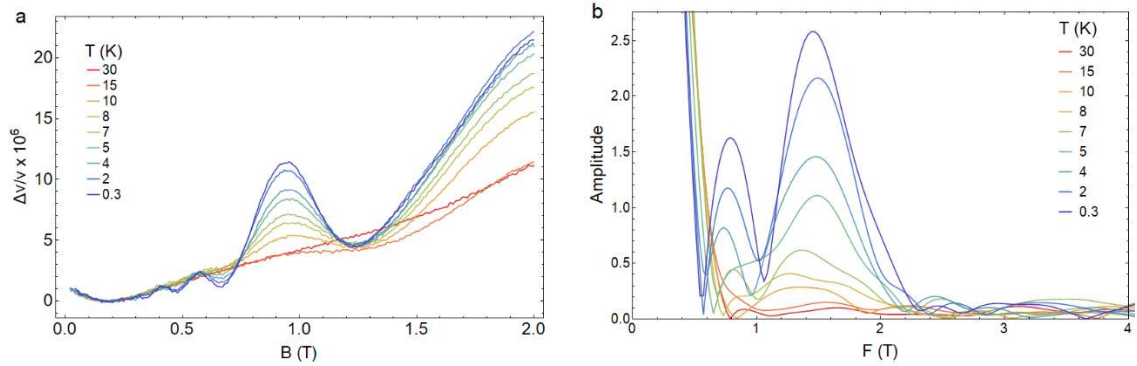

*Fig. S6 Change in the c-axis sound velocity with magnetic field applied along the [100] direction. a. The raw data measured at 802 MHz. Oscillations can be seen as low as 0.2 T, indicating high sample quality. b. Fourier transform of the data from the left panel between 0.24 and 2 Tesla. The large upturn at low frequency is due to the background, which has not been removed.*

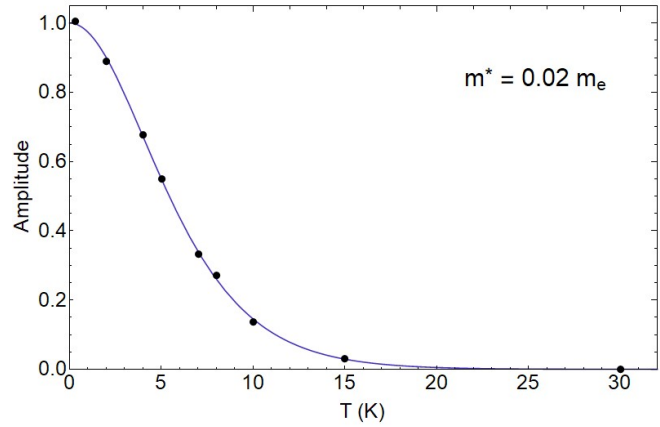

*Fig. S7 Lifshitz-Kosevich mass plot. The amplitude is defined as the maximum of Δv/v at approximately 0.9 Tesla minus the minimum at approximately 0.7 Tesla. The fit is to the standard LK expression, $A(T) = X/\sinh X$, where $X = \frac{2\pi^2 k_B T}{\hbar \omega_c}$, and $\omega_c = \frac{eB}{m^*}$ is the cyclotron frequency.*

Fig. S7 shows the quantum oscillation amplitude of the 1.4 T frequency as a function of temperature, yielding a cyclotron effective mass of $m^* = 0.02\ m_e$. The frequency and mass combine to give a Fermi velocity of $3.8\times10^5$ m/s – nearly a factor of 2 higher than what we measure at the W2 Weyl point in the infrared spectroscopy.

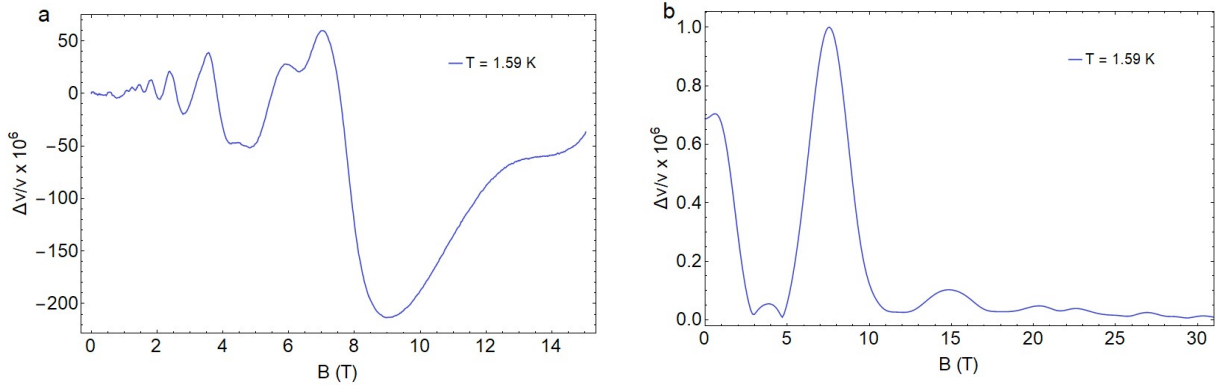

*Fig. S8 Change in the c-axis sound velocity with magnetic field applied along the [001] direction. a. Raw oscillatory data taken at the same ultrasonic frequency and with the same method as in Fig. S6 but for field along the c axis. b. Fourier transform of the data in a), showing a clear peak at F = 7.5 T, in agreement with the Fermi surface area of the W1 pocket measured previously (19).*

Figure S8 shows quantum oscillations in the sound velocity measured for field along the c axis. The main frequency of approximately 7.5 Tesla is consistent with what has been identified in other studies as belonging to the Fermi surface around the W1 Weyl node *(19,22)*.